\documentclass[10pt, conference, letterpaper]{IEEEtran}
\IEEEoverridecommandlockouts
\usepackage{graphicx}                  
\usepackage{booktabs}
\usepackage{amsmath, amsthm}

\usepackage{mdframed}
\usepackage{enumitem}
\usepackage{algorithm}

\usepackage{algpseudocode}
\usepackage{siunitx}
\usepackage{longtable}
\usepackage{array}
\usepackage{amsmath}
\usepackage{amssymb}
\usepackage{amsfonts}
\usepackage{tikz}
\usepackage{graphicx}
\usetikzlibrary{
    positioning,
    arrows.meta,
    calc
}
\usepackage{tabularx}
\usepackage{placeins}
\usepackage{dblfloatfix}
\usepackage{makecell}
\usepackage{adjustbox}
\usepackage{xcolor}
\usepackage{xurl}
\usepackage[
    colorlinks=true,
    linkcolor=blue,
    citecolor=red,
    urlcolor=magenta
]{hyperref}
\usepackage{balance}

\newcommand{\fixed}{\checkmark}
\newcommand{\varied}{\textbf{$\star$}}

\usetikzlibrary{arrows.meta,positioning}
\begin{document}

\date{}

\sloppy

\title{\huge Beyond Model Scaling: Rethinking Sequence Construction for Transformer-based Network Intrusion Detection\thanks{This work has been submitted to the IEEE for possible publication. Copyright may be transferred without notice, after which this version may no longer be accessible.}}
\author{\IEEEauthorblockN{Zeeshan Zulkifl Shah, Muhammad Ikram, Dali Kaafar, and Yipeng Zhou}
\IEEEauthorblockA{Macquarie University, Sydney, Australia\\
\{zeeshanzulkifl.shah,muhammad.ikram,dali.kaafar,yipeng.zhou\}@mq.edu.au}}

\maketitle

\begin{abstract}

Transformer-based network intrusion detection systems (NIDS) have shown strong potential for detecting multi-stage cyberattacks, yet most methods construct input sequences using fixed temporalization schemes rather than treating sequence construction as a systems design decision. In flow-based intrusion detection, timestamp-based windowing often fragments related attack activities while merging unrelated communications, reducing the behavioral context available to transformers. We present a unified pipeline abstraction that decomposes transformer-based NIDS into four design dimensions---traffic representation, temporalization, transformer backbone, and prediction aggregation---and propose \emph{session-coherent temporalization}, which constructs sequences from communication sessions instead of fixed time windows. Experiments on three complementary intrusion detection datasets show that temporalization is the dominant factor governing transformer performance: \emph{session-coherent temporalization} more than doubles Macro-F1 over timestamp-based windowing, enabling lightweight transformers to achieve up to 0.95 Macro-F1 at substantially higher inference throughput than
pretrained models. 
We further vary the transformer size from 0.65M to 3.38M parameters and find that larger transformer models provide no measurable accuracy improvement despite substantially higher computational cost, while zero-shot cross-dataset evaluation shows only modest 
gains under domain shift. These results indicate that preserving communication-session semantics, rather than scaling transformer capacity, is the key to accurate and efficient transformer-based intrusion detection.

\end{abstract}

\section{Introduction}
\label{sec:intro}
The widespread adoption of end-to-end encryption has fundamentally changed the operating environment of modern network intrusion detection systems (NIDS). As payload inspection becomes increasingly impractical, operational NIDS must rely primarily on flow-level telemetry, where attacks are manifested not by individual flows but through coordinated sequences of network interactions, including reconnaissance, exploitation, lateral movement, and command-and-control communication~\cite{hutchins2011killchain}. Consequently, effective intrusion detection increasingly depends on preserving these temporal relationships while simultaneously satisfying the latency and throughput requirements of high-speed operational networks.

Temporal dependencies provide essential evidence for understanding attack progression~\cite{shen2019attack2vec}, yet existing NIDS often fail to preserve them. Signature-based systems such as Suricata~\cite{suricata_doc} and Snort~\cite{snort_docs,roesch1999snort} rely on manually crafted rules and consequently struggle to detect previously unseen attacks~\cite{sommer2010outside}. Learning-based approaches improve generalization by modeling flow-level telemetry directly~\cite{buczak2016survey,chou2022csur}, while recent networking research has explored sequence-aware traffic representations for encrypted traffic analysis and lightweight malicious-traffic classification at the network edge~\cite{liu2019fsnet,diallo2021acid}. Nevertheless, many existing NIDS either classify flows independently or rely on fixed sequence-construction heuristics that fragment related attack stages while grouping unrelated communications into the same context. As encrypted traffic continues to dominate Internet-scale deployments~\cite{felt2017https,google_https}, preserving communication semantics has become a fundamental systems challenge rather than merely a machine learning problem.

Transformer architectures naturally model long-range dependencies through self-attention~\cite{vaswani2017attention}, making them attractive for flow-based intrusion detection~\cite{manocchio2024flowtransformer,han2023gtid}. Deployment-oriented networking systems have similarly investigated sequence-aware intrusion detection with execution inside programmable networks~\cite{zhao2024rids}. Unlike natural language, however, network traffic has no intrinsic sequential representation. Before transformer inference, asynchronous flows must first be organized into input sequences, a process we refer to as \emph{temporalization}~\cite{corsini2021evaluation}. Temporalization is therefore not merely a preprocessing step; it is a fundamental systems design decision that determines which communication dependencies are preserved for learning and which are irreversibly discarded.

Existing transformer-based NIDS~\cite{manocchio2024flowtransformer,martin2026transformer,han2023gtid} typically construct sequences using fixed timestamp windows or simple batching heuristics. Although computationally convenient, these strategies frequently separate related attack stages across different sequences while simultaneously merging unrelated communications into the same context, producing behaviorally incoherent inputs (see \S\ref{sec:session_temporalization}). Consequently, transformers are forced to learn from distorted communication context, limiting detection performance irrespective of model capacity. We argue that temporalization should be treated as a first-class systems design dimension rather than a fixed implementation detail. More fundamentally, we hypothesize that the effectiveness of transformer-based NIDS is governed more by sequence construction than by continually increasing transformer capacity.

Motivated by this observation, we introduce \emph{session-coherent temporalization}, a communication-session-aware sequence construction strategy that organizes flows according to communication sessions instead of arbitrary timestamp windows. By preserving communication semantics across related interactions, session-coherent temporalization enables transformers to model within-session attack behavior more coherently than under timestamp-based windowing.

More broadly, we present a \emph{unified transformer-NIDS design framework} (see \S\ref{sec:framework}) that decomposes the detection pipeline into four independent systems design dimensions: \emph{traffic representation}, \emph{temporalization}, \emph{transformer backbone}, and \emph{prediction aggregation}. This modular abstraction enables controlled evaluation of each design dimension in isolation while eliminating confounding factors that have complicated comparisons among existing transformer-based NIDS.

Using this framework, we systematically evaluate transformer-based NIDS across three complementary flow-based intrusion detection datasets: UNSW-NB15~\cite{moustafa2015unsw}, CICIDS2018~\cite{sharafaldin2018toward}, and MQCIDS~\cite{shah2025mqcids}. Our evaluation spans multiple traffic representations, temporalization strategies, transformer capacities, aggregation mechanisms, and context lengths while jointly considering detection effectiveness and deployment performance. The results demonstrate that temporalization, rather than transformer scale, is the dominant systems design choice governing transformer-based intrusion detection. More broadly, they advocate a pipeline-centric view of transformer-based NIDS, where engineering behavior-preserving sequence construction is more effective than continually scaling model architectures.

\noindent\textbf{Contributions.}
Our main contributions are as follows:

\begin{itemize}[leftmargin=*]

\item \textbf{Systems insight into transformer-based NIDS.}
We demonstrate that \emph{temporalization} is the dominant systems design dimension governing transformer-based intrusion detection. Treating sequence construction as fixed preprocessing fundamentally confounds the evaluation of transformer architectures, whereas preserving communication-session semantics consistently contributes larger gains than increasing model capacity.

\item \textbf{Session-coherent temporalization.}
We propose \emph{session-coherent temporalization}, a communication-session-aware sequence construction strategy that organizes flows according to communication sessions rather than arbitrary timestamp windows. The proposed method preserves within-session behavioral continuity, enabling transformers to learn from communication-consistent contexts while reducing fragmentation and cross-session contamination.

\item \textbf{Comprehensive deployment-oriented evaluation.} Through extensive experiments on UNSW-NB15, CICIDS2018, and MQCIDS, we show that session-coherent temporalization more than doubles Macro-F1 over conventional timestamp-based windowing while enabling compact transformers to achieve up to \textit{0.95 Macro-F1} and sustain inference throughput approaching \textit{$10^6$ processed flow records per second}. We further show that scaling transformer capacity yields negligible benefit, whereas cross-dataset experiments indicate that domain shift remains the principal barrier to robust zero-shot deployment.

\end{itemize}
For reproducibility, we release an open-source artifact containing implementation details, experimental evidence, and supplementary materials~\cite{artifact2027}.

\section{Background and Related Work}
\label{sec:background}
Transformer-based NIDS depend on both the learning architecture and how network traffic is transformed into sequential input. This section formalizes temporalization, reviews existing approaches, and motivates sequence construction as an independent systems design dimension.

\subsection{Flow-based Intrusion Detection and Temporalization}

Modern NIDS increasingly rely on \emph{flow-level telemetry} because it scales well to high-speed networks and remains applicable when payload inspection is limited by encryption.  

Unlike natural language, network traffic has no intrinsic sequential structure. Flow records generated by independent communications must therefore be transformed into ordered sequences before sequential learning models can be applied. We refer to this transformation as \emph{temporalization}.

Formally, let $X=(x_1,x_2,\ldots,x_N)$ denote an ordered stream of flow records, where each flow $x_i\in\mathbb{R}^d$ is represented by a feature vector. A temporalization function

$\tau:X\rightarrow\{W_1,\ldots,W_M\}$

maps the flow stream into sequential windows

$W_t=(x_t,\ldots,x_{t+k-1}),$

which are processed by a classifier

$g_\theta:W_t\rightarrow\hat{y}_t.$

Temporalization is independent of the learning architecture; identical transformer models may therefore observe entirely different behavioral contexts depending on how sequences are constructed.

Existing systems typically instantiate $\tau$ using chronological flow sequences, fixed timestamp windows, or dataset-specific batching heuristics~\cite{liu2019fsnet, han2023gtid,manocchio2024flowtransformer}. Although efficient, these strategies may fragment related attack activities while grouping unrelated communications into the same sequence, limiting the behavioral information available for inference.

\subsection{Related Work and Research Gap}

Traditional NIDS such as Snort and Suricata rely on manually engineered signatures and therefore struggle against previously unseen attacks~\cite{roesch1999snort, suricata_doc, snort_docs,sommer2010outside}. Learning-based methods improve generalization by modeling flow-level telemetry directly~\cite{buczak2016survey,chou2022csur}.

Within the networking community, FS-Net represents encrypted traffic as flow sequences, ACID targets lightweight malicious-traffic classification at the network edge, and RIDS co-designs a sequential detector with programmable switches for high-speed deployment~\cite{liu2019fsnet,diallo2021acid,zhao2024rids}. Transformer-based traffic models use self-attention and pretraining to learn contextual flow or datagram representations~\cite{vaswani2017attention, manocchio2024flowtransformer, han2023gtid, lin2022etbert}, while domain-specific pretraining has also been explored through CTI-BERT~\cite{park_you_2023_ctibert}.

Recent research has additionally moved beyond static offline detection. AOC-IDS combines contrastive representation learning with pseudo-label-driven online updates to adapt the detector as traffic behavior and attack strategies evolve~\cite{zhang2024aocids}.

Despite these advances, existing evaluations remain largely \emph{model-centric}: transformer architectures are compared under a fixed temporalization strategy inherited from the dataset or preprocessing pipeline. Consequently, the effects of sequence construction are confounded with model capacity, making it unclear whether observed improvements arise from stronger architectures or from the behavioral information preserved in the input sequences.

To address this limitation, we introduce a unified transformer-based NIDS framework (cf.~\S\ref{sec:framework}) that independently varies traffic representation, temporalization, transformer backbone, and prediction aggregation. This decomposition enables controlled evaluation of each design dimension and demonstrates that preserving behavioral coherence during sequence construction consistently contributes larger performance gains than increasing transformer capacity (see \S\ref{sec:results}).

\section{System Design}
\label{sec:framework}
Rather than focusing on another transformer architecture, this work formulates transformer-based intrusion detection as a pipeline of independently configurable design dimensions.

\begin{figure*}[!ht]
\centering

\resizebox{\textwidth}{!}{
\begin{tikzpicture}[
    font=\sffamily\scriptsize,
    node distance=3.2mm,
    stage/.style={
        draw=blue!55!black,
        fill=blue!5,
        rounded corners=2pt,
        line width=0.65pt,
        minimum width=2.25cm,
        minimum height=2.55cm,
        text width=1.95cm,
        align=center,
        inner xsep=0.5mm,
        inner ysep=0.3mm
    },
    keyStage/.style={
        stage,
        draw=orange!85!black,
        fill=orange!14,
        line width=0.9pt,
        minimum width=2.55cm,
        text width=2.25cm
    },
    connector/.style={
        -{Latex[length=2.2mm,width=1.5mm]},
        line width=0.85pt,
        draw=black!85,
        shorten <=1pt,
        shorten >=1pt
    },
    iconLine/.style={
        draw=black!58,
        line width=0.50pt
    }
]

\node[stage] (data) {
    
    \textbf{Data Sources}\\[-0.2mm]
    UNSW-NB15\\
    CICIDS2018\\
    MQCIDS
};

\node[stage, right=of data] (harm) {
    
    \textbf{Harmonization}\\[-0.2mm]
    Schema alignment\\
    Feature derivation
};

\node[stage, right=of harm] (control) {
    
    \textbf{Input Control\\
    $(\psi)$}\\[-0.2mm]
    ID sanitization\\
    Train-only\\
    normalization
};

\node[stage, right=of control] (repr) {
    
    \textbf{Representation $(\phi)$}\\[-0.2mm]
    Per-flow\\
    field tokens\\
    Serialized text
};

\node[keyStage, right=of repr] (temp) {
    
    \textbf{Temporalization $(\tau)$}\\[-0.2mm]
    \textbf{Sliding windows}\\
    \textbf{vs.}\\
    \textbf{session-coherent}\\
    \textbf{windows}
};

\node[stage, right=of temp] (backbone) {
    
    \textbf{Transformer Backbone\\ $(f_\theta)$}\\[-0.2mm]
    Shallow models\\
    Pretrained\\
    models
};

\node[stage, right=of backbone] (aggregation) {
    
    \textbf{Aggregation\\
    $(\mathrm{Agg})$}\\[-0.2mm]
    Mean pooling\\
    Attention pooling\\
    CLS / last token
};

\node[stage, right=of aggregation] (prediction) {
    
    \textbf{Prediction\\ 
    $(\hat{y})$}\\[-0.2mm]
    Intrusion label
};

\draw[connector] (data.east) -- (harm.west);
\draw[connector] (harm.east) -- (control.west);
\draw[connector] (control.east) -- (repr.west);
\draw[connector] (repr.east) -- (temp.west);
\draw[connector] (temp.east) -- (backbone.west);
\draw[connector] (backbone.east) -- (aggregation.west);
\draw[connector] (aggregation.east) -- (prediction.west);

\begin{scope}[shift={($(data.south)+(0,0.32)$)}, scale=0.82, transform shape]
    \draw[iconLine] (0,0.14) ellipse (0.28 and 0.08);
    \draw[iconLine] (-0.28,0.14) -- (-0.28,-0.15);
    \draw[iconLine] ( 0.28,0.14) -- ( 0.28,-0.15);
    \draw[iconLine]
        (-0.28,-0.15)
        arc[start angle=180,end angle=360,
            x radius=0.28,y radius=0.08];
    \draw[iconLine]
        (-0.28,-0.01)
        arc[start angle=180,end angle=360,
            x radius=0.28,y radius=0.08];
\end{scope}

\begin{scope}[shift={($(harm.south east)+(-1.10,0.24)$)}]
    \draw[iconLine,<->] (-0.38,0) -- (0.38,0);
    \draw[iconLine] (-0.16,-0.14) -- (-0.16,0.14);
    \draw[iconLine] ( 0.16,-0.14) -- ( 0.16,0.14);
\end{scope}

\begin{scope}[shift={($(control.south east)+(-1.10,0.24)$)}]
    \draw[iconLine,rounded corners=1pt]
        (-0.36,-0.13) rectangle (-0.10,0.13);
    \draw[iconLine,rounded corners=1pt]
        (0.08,-0.13) rectangle (0.34,0.13);
    \draw[iconLine] (-0.23,0.13) -- (-0.23,0.28);
    \draw[iconLine] ( 0.21,0.13) -- ( 0.21,0.28);
\end{scope}

\begin{scope}[shift={($(repr.south east)+(-1.10,0.24)$)}]
    \foreach \x/\lab in {
        -0.43/$x_1$,
        -0.14/$x_2$,
         0.15/$\cdots$,
         0.44/$x_d$
    }{
        \node[
            draw=black!65,
            rounded corners=0.6pt,
            minimum width=0.23cm,
            minimum height=0.21cm,
            inner sep=0.5pt,
            font=\tiny
        ] at (\x,0) {\lab};
    }
\end{scope}

\begin{scope}[shift={($(temp.south)+(0,0.24)$)}, scale=0.70, transform shape]
    
    \draw[iconLine,rounded corners=1pt]
        (-0.55,-0.13) rectangle (-0.27,0.11);
    \draw[iconLine,rounded corners=1pt]
        (-0.38,-0.03) rectangle (-0.10,0.21);
    \draw[iconLine,rounded corners=1pt]
        (-0.21,0.07) rectangle (0.07,0.31);

    \draw[iconLine] (0.30,0.00) circle (0.07);
    \draw[iconLine] (0.49,0.00) circle (0.07);
    \draw[iconLine] (0.68,0.00) circle (0.07);
    \draw[iconLine] (0.37,0.00) -- (0.42,0.00);
    \draw[iconLine] (0.56,0.00) -- (0.61,0.00);
\end{scope}

\begin{scope}[shift={($(backbone.south)+(0,0.24)$)}, scale=0.82, transform shape]
    \foreach \y in {-0.12,0.03,0.18}{
        \foreach \x in {-0.34,-0.11,0.11,0.34}{
            \fill[black!60] (\x,\y) circle (1.05pt);
        }
    }
    \draw[iconLine] (-0.34,-0.12) -- (0.34,0.18);
    \draw[iconLine] (-0.34,0.18) -- (0.34,-0.12);
    \draw[iconLine] (-0.11,-0.12) -- (0.11,0.18);
\end{scope}

\begin{scope}[shift={($(aggregation.south east)+(-1.10,0.24)$)}]
    \draw[iconLine] (-0.34,-0.05) circle (0.065);
    \draw[iconLine] (0,-0.05) circle (0.065);
    \draw[iconLine] (0.34,-0.05) circle (0.065);
    \draw[iconLine] (0,0.24) circle (0.075);
    \draw[iconLine] (-0.28,0.00) -- (-0.05,0.19);
    \draw[iconLine] (0,0.02) -- (0,0.17);
    \draw[iconLine] (0.28,0.00) -- (0.05,0.19);
\end{scope}

\begin{scope}[shift={($(prediction.south)+(0,0.34)$)}, scale=0.78, transform shape]
    \draw[iconLine] (0,0.04) circle (0.24);
    \draw[iconLine] (0,0.04) circle (0.14);
    \draw[iconLine] (0,0.04) circle (0.045);
    \draw[
        -{Latex[length=1.5mm,width=1.0mm]},
        line width=0.65pt,
        draw=black!70
    ]
        (-0.30,-0.22) -- (-0.02,0.02);
\end{scope}

\node[
    font=\sffamily\tiny\bfseries,
    text=orange!70!black,
    above=1.2mm of temp
] {
    PRIMARY DESIGN DIMENSION
};

\end{tikzpicture}
}
\vspace{-1.15em}
\caption{\scriptsize
Overview of the proposed transformer-based NIDS framework. The pipeline separates input control from four independently
configurable design dimensions: traffic representation, temporalization, transformer backbone, and prediction aggregation.}

\label{fig:pipeline_overview}
\vspace{-2.15em}
\end{figure*}
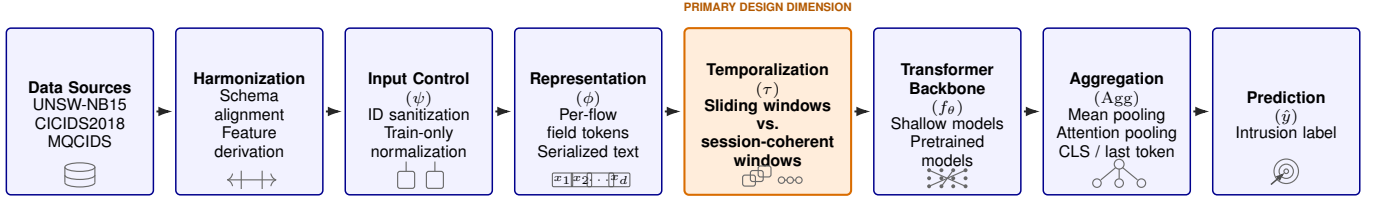

\subsection{Pipeline Abstraction}
Given a stream of network flows $X$, the proposed framework computes
\[
\hat{y}
=
\mathrm{Agg}
\circ
f_\theta
\circ
\tau
\circ
\phi
\circ
\psi(X),
\]
where:
\begin{itemize}[leftmargin=*]
\item Input control $\psi$, which removes identifier-based features and standardizes numeric attributes,
\item Representation $\phi$, which converts flow records into transformer tokens,
\item Temporalization $\tau$, which constructs sequential contexts,
\item Transformer backbone $f_\theta$, which learns contextual relationships within each sequence, and
\item Aggregation $\mathrm{Agg}$, which maps token embeddings to a sequence prediction.
\end{itemize}

Fig.~\ref{fig:pipeline_overview} illustrates the proposed abstraction. Unlike prior transformer-based NIDS that primarily compare model architectures under a fixed preprocessing pipeline, the proposed framework independently controls traffic representation, temporalization, transformer backbone, and prediction aggregation. This enables controlled evaluation of each design dimension while isolating the contribution of temporalization from that of the learning architecture.

\subsection{Independent Design Dimensions}

Our framework exposes four independently configurable design dimensions: \textit{(i)} traffic representation, \textit{(ii)} temporalization, \textit{(iii)} transformer backbone, and \textit{(iv)} prediction aggregation. From the moment a network flow enters the pipeline until an intrusion decision is made, each stage contributes a distinct design choice. Traffic representation defines how flows are encoded into transformer inputs, temporalization determines which flows are observed together as behavioral context, the transformer backbone learns contextual relationships within that context, and prediction aggregation converts contextual representations into the final intrusion prediction. By explicitly separating these stages, the framework enables controlled evaluation of each design dimension while eliminating the confounding present in prior transformer-based NIDS studies~\cite{han2023gtid,manocchio2024flowtransformer}, where sequence construction is typically inherited from a fixed preprocessing pipeline. 

Although the framework supports independent variation of all four dimensions, this work focuses on temporalization because it is the only stage that determines the behavioral information available before learning begins. Temporalization constitutes the primary design dimension in this work; the remaining dimensions are either fixed or varied through controlled ablations to quantify their independent and joint effects relative to temporalization.

\subsection{Session-Coherent Temporalization}
\label{sec:session_temporalization}

Preserving communication structure during temporalization is critical because transformer self-attention operates only within each constructed sequence; discarded relationships beforehand cannot be recovered by downstream learning, regardless of model capacity or architectural sophistication.

To characterize structural preservation, let session $j$ contain
$n_j$ chronologically ordered flows
$(x_{j,1},\ldots,x_{j,n_j})$, with $p<q$ indexing flow positions.
Let $\mathcal{W}_A$ denote the set of windows produced by a particular temporalization
strategy $A$ with window length $k$, and let $\mathbb{I}[\cdot]$
denote the indicator function. Related-flow retention measures the
fraction of emittable within-session flow pairs that co-occur in at
least one window:
{\scriptsize \[
R_{j,\mathrm{emit}}^{(k)}
=
\left\{
(x_{j,p},x_{j,q})
\mid
n_j\geq k,\;
1\leq p<q\leq n_j,\;
q-p<k
\right\},
\]}
with 
{\scriptsize \[
\operatorname{Ret}(A)
=
\frac{
\sum_j
\sum_{(x_{j,p},x_{j,q})\in R_{j,\mathrm{emit}}^{(k)}}
\mathbb{I}
\left[
\exists W\in\mathcal{W}_A:
\{x_{j,p},x_{j,q}\}\subseteq W
\right]
}{
\sum_j |R_{j,\mathrm{emit}}^{(k)}|
}.
\]}

The complementary fragmentation rate is 
$\operatorname{Frag}(A)=1-\operatorname{Ret}(A)$ while window purity measures the fraction of flows belonging to the dominant communication session within each constructed window,
\[
\operatorname{Purity}(A)
=
\frac{1}{|\mathcal{W}_A|}
\sum_{W\in\mathcal{W}_A}
\frac{\max_j c_j(W)}{|W|},
\]
where $c_j(W)$ denotes the number of flows from session $j$ contained in window $W$. Higher retention and purity, together with lower fragmentation, indicate that more related flows are grouped into the same windows, implying a better preservation of behavioral structure.

Table~\ref{tab:session_temporalization} shows that,
across the three datasets, session-coherent temporalization
increases average related-flow retention from 0.096 to 0.752
and purity from 0.190 to 1.000, while reducing fragmentation
from 0.904 to 0.248.

To address the limitations of timestamp-based construction, we generate windows independently within each communication session rather than over the globally ordered traffic stream. Let $b(x_i)$ denote the communication
session associated with flow $x_i$, and let
\[
S_j=(x_{j,1},x_{j,2},\ldots,x_{j,n_j})
\]
denote the chronologically ordered sequence of $n_j$ flows belonging to
session $j$. Given window length $k$ and stride $s$, session $j$ emits
\[
M_j=
\max\left\{
0,
\left\lfloor\frac{n_j-k}{s}\right\rfloor+1
\right\}
\]
windows. For $m=0,1,\ldots,M_j-1$, the $m$-th window is
\[
W_m^{(j)}
=
\left(
x_{j,1+ms},
x_{j,2+ms},
\ldots,
x_{j,k+ms}
\right).
\]

Unlike timestamp-based temporalization, every emitted window contains flows from exactly one communication session, preserving behavioral continuity while preventing cross-session contamination.

Algorithm~\ref{alg:session_temporalization} summarizes the proposed procedure. Flows are first grouped by session identifier and ordered chronologically. Sliding windows are then generated independently within each session using window length $k$ and stride $s$. Sessions shorter than $k$ are omitted to avoid introducing artificial padded contexts and to ensure that every emitted window corresponds to observed communication behavior.

The total number of session-coherent windows is
$M_{\mathrm{sess}}=\sum_j M_j$. By comparison, sliding over the globally
ordered stream of $N=\sum_j n_j$ flows produces
\[
M_{\mathrm{slide}}
=
\max\left\{
0,
\left\lfloor\frac{N-k}{s}\right\rfloor+1
\right\}.
\]
For the matched overlapping-window setting $1\leq s\leq k$,
$M_{\mathrm{sess}}\leq M_{\mathrm{slide}}$, because session boundaries
prevent windows from spanning unrelated communications and sessions
shorter than $k$ emit no windows. Moreover, each flow appears in at most
$\lceil k/s\rceil$ windows, which is at most two under
$(k,s)=(32,16)$.

\begin{algorithm}[!ht]
\small
\caption{\small Session-Coherent Window Construction}
\label{alg:session_temporalization}
\begin{algorithmic}[1]
\Require Flows $\mathcal{X}$, window length $k$, stride $s$
\Ensure Window set $\mathcal{W}$
\State $\mathcal{W}\gets\emptyset$
\ForAll{$S_j \in \Call{GroupAndSortBySession}{\mathcal{X}}$}
    \If{$|S_j|<k$}
        \State \textbf{continue}
    \EndIf
    \For{$a\gets1$ \textbf{to} $|S_j|-k+1$ \textbf{step} $s$}
        \State $\mathcal{W}\gets\mathcal{W}\cup
        \{S_j[a:a+k-1]\}$
    \EndFor
\EndFor
\State \Return $\mathcal{W}$
\end{algorithmic}
\end{algorithm}
\vspace{-1em}
\begin{table}[!ht]
\centering
\scriptsize
\caption{\small Structural preservation under matched window length and stride $(k=32,s=16)$. Higher retention and purity, together with lower fragmentation, indicate stronger preservation of communication structure.}
\label{tab:session_temporalization}
\setlength{\tabcolsep}{2.0pt}
\begin{tabular}{lcccccc}
\toprule
& \multicolumn{2}{c}{Retention $\uparrow$} &
  \multicolumn{2}{c}{Fragmentation $\downarrow$} &
  \multicolumn{2}{c}{Purity $\uparrow$} \\
\cmidrule(lr){2-3}\cmidrule(lr){4-5}\cmidrule(lr){6-7}
Dataset & Sliding & Session & Sliding & Session & Sliding & Session \\
\midrule
UNSW   & 0.114 & \textbf{0.756} & 0.886 & \textbf{0.244} & 0.286 & \textbf{1.000} \\
CICIDS & 0.027 & \textbf{0.742} & 0.973 & \textbf{0.258} & 0.049 & \textbf{1.000} \\
MQCIDS & 0.148 & \textbf{0.757} & 0.852 & \textbf{0.243} & 0.236 & \textbf{1.000} \\
\midrule
Avg.   & 0.096 & \textbf{0.752} & 0.904 & \textbf{0.248} & 0.190 & \textbf{1.000} \\
\bottomrule
\end{tabular}
\vspace{-0.8em}
\end{table}

Dataset-specific session reconstruction and window configurations are described in $\S$\ref{sec:experimental_setup}.

\subsection{Transformer Learning}

Our framework decouples behavioral sequence construction from transformer learning. Let $W=(x_1,\ldots,x_n)$ denote an emitted flow window produced by the temporalization stage. Each flow $x_i$ is transformed into a model-specific token representation $\phi(x_i)$, yielding the input token sequence: $T=(\phi(x_1),\ldots,\phi(x_n))=(t_1,\ldots,t_n)$. The transformer computes contextualized embeddings $H=f_\theta(T)$ and the aggregation operator produces the final prediction $\hat{y}=\mathrm{Agg}(H)$ following the standard transformer formulation~\cite{vaswani2017attention}.

The proposed framework is intentionally model-agnostic, allowing different
transformer backbones to operate on identical temporalized inputs.
Throughout this work, lightweight transformers trained from scratch and
representative pretrained encoder- and decoder-based models are evaluated
under controlled temporalization settings, as detailed in
$\S$\ref{sec:experimental_setup}. This design enables systematic
comparison of temporalization strategies and model capacity while keeping the
remaining pipeline components fixed within each experiment. Implementation
details are provided in our artifact repository~\cite{artifact2027}.

\subsection{Computational Complexity and Deployment Considerations}

The proposed temporalization targets continuous online monitoring, consistent with high-throughput IDS/IPS and in-network systems that prioritize processing efficiency~\cite{zhao2020pigasus,zhao2024rids}. Conventional sliding-window temporalization performs a single pass over incoming flows, whereas session-coherent temporalization additionally maintains lightweight state for active sessions before emitting windows independently within each session.

Under bounded active-session state, both approaches execute in linear time in the number of processed flows, while session-coherent temporalization requires additional state proportional to the number of concurrently active sessions. The extra computation consists of maintaining a hash table of active sessions and appending each incoming flow to its corresponding session buffer, both of which have expected constant-time cost. Consequently, the proposed approach preserves the overall linear preprocessing complexity of the pipeline.

To quantify practical overhead, we performed a controlled streaming microbenchmark in which all pipeline components except
temporalization were held constant; complete measurements are provided in the artifact~\cite{artifact2027}. Both methods processed the same timestamp-ordered flow stream using identical window length, stride, rolling-window logic, and hardware configuration. Session identifiers were precomputed so that the benchmark isolated only the cost of session-aware window construction and state maintenance.

Compared with timestamp-based global sliding-window construction,
session-coherent temporalization required \(2.25\times\) the processing
time on average. The main and complete-session robustness benchmarks
yielded runtime ratios of \(2.1669\times\) and \(2.3304\times\),
respectively; their arithmetic mean is
\((2.1669+2.3304)/2=2.2487\approx2.25\times\).

Despite this overhead, session-coherent temporalization sustained \(1.04\) million input flows/s while emitting \(54\%\) fewer windows for downstream inference. The benchmark isolates temporalization rather than end-to-end throughput, but the lower window count reduces subsequent inference calls. The window-construction stage therefore remains
operationally feasible while providing the detection gains reported in $\S$\ref{sec:results}.

\section{Evaluation Methodology}
\label{sec:evaluation}

Having introduced the pipeline abstraction, we next evaluate its design dimensions through a series of controlled experiments. Each research question varies only the pipeline component or component set required to answer it,
while all remaining components are held fixed.

\subsection{Research Questions}
The evaluation is organized around five research questions that progressively investigate the contribution of temporalization, model capacity, deployment efficiency, and cross-dataset robustness within the proposed pipeline abstraction.

\begin{itemize}[leftmargin=*]

\item \textbf{RQ1 (Temporalization).}
How much does session-coherent temporalization improve transformer-based NIDS compared with conventional timestamp-based windowing?

\item \textbf{RQ2 (Model Capacity).}
Once behavioral context is preserved, how much additional benefit is gained by increasing transformer capacity?

\item \textbf{RQ3 (Pipeline Optimization versus Foundation Models).}
Does improving sequence construction provide greater benefit than scaling to large pretrained transformer models?

\item \textbf{RQ4 (Deployment Trade-offs).}
How do different temporalization strategies and transformer architectures trade detection effectiveness against deployment efficiency?

\item \textbf{RQ5 (Cross-Dataset Generalization).}
Does session-coherent temporalization improve zero-shot transfer across heterogeneous network environments, or do dataset-specific traffic and label semantics remain the dominant limitation?

\end{itemize}

\subsection{Experimental Setup}
\label{sec:experimental_setup}

The pipeline abstraction introduced in $\S$\ref{sec:framework} defines our experimental methodology. Each RQ's primary comparison varies only the relevant pipeline dimension(s) while keeping all others fixed (Table~\ref{tab:rq_mapping}), enabling controlled attribution of performance differences to individual design choices.

\begin{table}[t]
\centering

\caption{Primary Pipeline components varied ($\star$) or fixed ($\checkmark$) for RQs.} 
\vspace{-0.2cm}
\label{tab:rq_mapping}
\scriptsize
\setlength{\tabcolsep}{2pt}
\renewcommand{\arraystretch}{1.15}
\begin{tabular}{c|c c c c}
\hline
\textbf{RQ} &
\textbf{Representation} &
\textbf{Temporalization} &
\textbf{Backbone} &
\textbf{Aggregation} \\
\hline
RQ1 & \fixed & \varied & \fixed & \fixed \\
RQ2 & \fixed & \fixed & \varied & \fixed \\
RQ3 & \fixed & \fixed & \varied & \fixed \\
RQ4 & \fixed & \varied & \varied & \fixed \\
RQ5 & \fixed & \varied & \fixed & \fixed \\
\hline
\end{tabular}
\vspace{-0.5cm}
\end{table}

\noindent\textbf{Datasets.} To evaluate the proposed framework under diverse operating conditions, we use three complementary flow-based intrusion detection datasets: UNSW-NB15, 
CICIDS2018, 
and MQCIDS. 
Collectively, they capture conventional enterprise traffic, contemporary cyber attacks, and deception-based operational environments. UNSW-NB15 and CICIDS2018 are controlled benchmark datasets containing scripted attack scenarios, whereas MQCIDS captures operational attacker telemetry through a deception infrastructure. For cross-dataset evaluation, dataset-specific features and labels are mapped to the common schema documented in the artifact~\cite{artifact2027}. Consequently, cross-dataset evaluation measures not only whether temporalization preserves behavioral context, but also whether learned traffic and label semantics transfer across heterogeneous network environments.

For UNSW-NB15 and CICIDS2018, communication sessions are reconstructed using the canonical five-tuple together with the inactivity timeout employed by the corresponding preprocessing pipeline. For MQCIDS, we directly use the Suricata-derived behavioral session identifiers generated during traffic collection. Unless otherwise stated, all experiments use the matched configuration $(k,s)=(32,16)$, while sensitivity analyses over $k\in\{8,16,32\}$ and multiple stride settings are provided in the artifact~\cite{artifact2027}. 

Each dataset is partitioned into 70\% training, 15\% validation, and 15\% testing splits before window construction, with communication sessions kept disjoint across partitions. Thus, no session contributes windows to more than one split. Feature harmonization, identifier removal, and normalization follow the preprocessing pipeline described in $\S$\ref{sec:framework}. Normalization statistics are measured only from the training split, preventing information leakage from validation and test data through preprocessing parameters (e.g., feature means and standard deviations).

\noindent\textbf{Transformer Models.}
To investigate the role of model capacity independently of sequence construction, we evaluate two complementary classes of transformer models.

\noindent\textit{Lightweight Transformers.} 
Compact transformer backbones are trained from scratch over a controlled architecture grid with hidden dimensions $d\in\{64,128,256\}$, depths $L\in\{1,2,4\}$, and attention heads $H\in\{2,4,8\}$. We refer to these from-scratch lightweight models as shallow transformer (ST) configurations. Since these models are not pretrained checkpoints, their names denote architectural operating points rather than externally released model families. Specifically, ST-B1, ST-B2, ST-B3, and ST-B4 denote representative shallow configurations selected from the capacity grid. Both encoder-style (ST-Enc) and decoder-style (ST-Dec) models are implemented in PyTorch~\cite{paszke2019pytorch} using standard transformer architecture~\cite{vaswani2017attention}.

\noindent\textit{Pretrained Foundation Models.}
ModernBERT~\cite{warner2025modernbert} and Qwen2.5-0.5B~\cite{qwen2025qwen25technicalreport} are used as the representative pretrained encoder and decoder models in the main paper, respectively. To demonstrate that the observed trends are not specific to these architectures, we additionally evaluate DeBERTa-v3~\cite{he2023debertav3} and the CTI-BERT checkpoint (\texttt{ibm-research/CTI-BERT})~\cite{park_you_2023_ctibert,ibmresearch_ctibert_hf}. Complete comparative and ablation results are provided in the artifact~\cite{artifact2027}. All pretrained models are implemented using the transformers library~\cite{wolf-etal-2020-transformers} and evaluated under identical representation, temporalization, and aggregation settings (cf. Table~\ref{tab:rq_mapping}) wherever applicable.

\noindent{\bf Evaluation Metrics.} The RQs require evaluation from three complementary perspectives: detection effectiveness, deployment efficiency, and robustness under domain shift.

\textit{Detection effectiveness} is primarily measured using Macro-F1, averaged over labels present in each evaluation split, with PR-AUC ~\cite{pr-auc} and accuracy reported as complementary metrics.

\textit{Deployment efficiency} is evaluated using processed-flow throughput together with inference latency, GPU memory consumption, CPU/GPU utilization, and model size, all of which characterize the practical cost of deployment.

Processed-flow throughput is defined as
\[
\mathrm{Throughput}
=
\frac{N_{\mathrm{eff}}}
{T_{\mathrm{infer}}}
\]
where $N_{\mathrm{eff}}$ counts flow positions processed by the model, including repeated positions in overlapping windows, and $T_{\mathrm{infer}}$ is the measured inference time. Unless otherwise stated, throughput is averaged over 50 random
batches, each run four times using the median runtime.

To quantify robustness under domain shift, we additionally report the transfer robustness gap (TRG),
\[
\mathrm{TRG}
=
\mathrm{Macro\mbox{-}F1}_{\mathrm{within}}
-
\mathrm{Macro\mbox{-}F1}_{\mathrm{cross}},
\]
where
$\mathrm{Macro\mbox{-}F1}_{\mathrm{within}}$
denotes the conventional within-dataset evaluation score and
$\mathrm{Macro\mbox{-}F1}_{\mathrm{cross}}$
denotes performance when evaluated on an unseen dataset.
Lower $TRG$ indicates stronger cross-domain transfer.

For RQ5, each directed source--target transfer setting is treated as one paired evaluation task because reversing the transfer direction changes both the training distribution and the independently trained source-domain model.
We compare Sliding and Session across the six directed tasks using an exact one-sided Wilcoxon signed-rank test, consistent with the directional hypothesis that session coherence improves transfer. We additionally report rank-biserial correlation as an effect-size measure.

\noindent\textbf{Controlled Evaluation.} Within each comparison, all components not marked as varied ($\star$) in Table~\ref{tab:rq_mapping} remain fixed ($\checkmark$).

\section{Experimental Results and Analysis}
\label{sec:results}
Next, we evaluate our framework introduced in $\S$\ref{sec:framework} by isolating  
the contribution of each design dimension. 
We begin with temporalization, since it determines

the behavioral context available to the transformer, before examining model capacity, prediction aggregation, deployment efficiency, and cross-dataset robustness.

\subsection{RQ1: Temporalization Dominates Detection Performance}
\label{subsec:finding1}

\noindent\textbf{Finding.} Session-coherent temporalization is the dominant design decision in the proposed pipeline. Compared with conventional timestamp-based windowing, it more than doubles Macro-F1 
while maintaining essentially identical processed-flow inference throughput. 

\begin{figure}[!th]
\centering
\vspace{-0.4cm}
\includegraphics[width=\columnwidth]{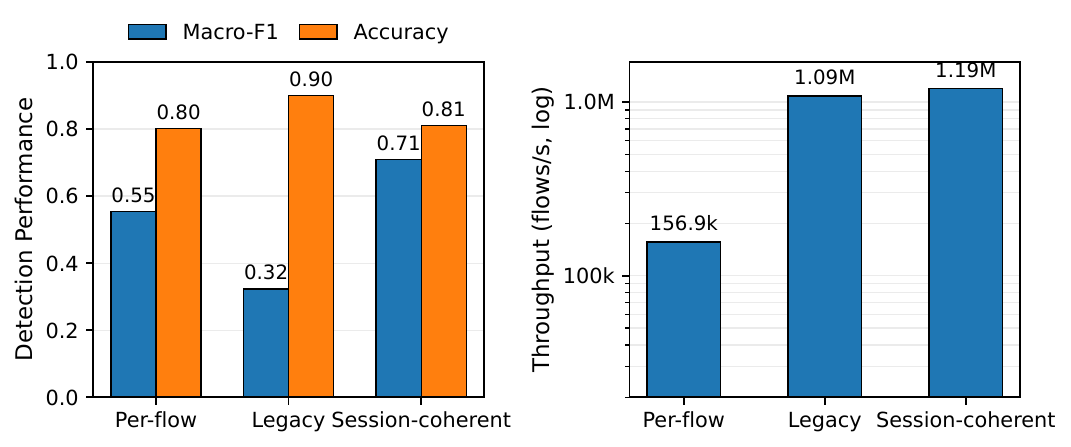}
\vspace{-2.5em}
\caption{\scriptsize Impact of temporalization under an identical shallow-transformer configuration $(d,L,H)=(128,4,4)$. \textit{Left:} detection performance for independent per-flow classification, legacy timestamp-based windowing, and the proposed session-coherent temporalization. \textit{Right:} processed-flow inference throughput. }

\label{fig:temporalization_results}
\vspace{-0.25em}
\end{figure}

Fig.~\ref{fig:temporalization_results} isolates the effect of temporalization by evaluating three sequence-construction strategies under an identical shallow-transformer configuration $(d,L,H)$ $=$ $(128,4,4)$. 
Because the learning architecture is unchanged, any observed performance difference can be attributed solely to how behavioral context is constructed before learning. Independent per-flow classification ignores relationships across communications, while timestamp-based windows preserve chronological order but frequently fragment attack progression or mix unrelated traffic. In contrast, session-coherent temporalization preserves communication-session semantics, enabling the transformer to reason over complete behavioral sequences rather than isolated flow observations.

The higher retention and purity in
Table~\ref{tab:session_temporalization} help explain
the detection gains by preserving within-session context.

Session-coherent temporalization more than doubles Macro-F1 while maintaining comparable processed-flow inference throughput. Because all three configurations employ the same transformer architecture, this improvement cannot be attributed to model capacity or additional computation. Instead, it arises from exposing the transformer to behaviorally coherent input sequences. This result identifies temporalization, rather than the learning architecture itself, as the dominant bottleneck in flow-based intrusion detection.

 \noindent\textbf{Representation Cannot Recover Lost Behavioral Context.} 
 To verify that the observed gains are not specific to a particular traffic encoding, we fix the temporalization strategy and transformer architecture while varying only the input representation. 
 Although the per-flow representation achieves the highest Macro-F1 (0.761), followed by field-token (0.580) and serialized-text (0.562), these differences are modest compared with the 2.2$\times$ improvement achieved by session-coherent temporalization (Fig.~\ref{fig:temporalization_results}). This demonstrates that traffic representation influences how behavioral information is encoded, whereas temporalization determines whether that information is available to the model. Consequently, once behavioral continuity is lost during sequence construction, no input representation alone can recover it.

\noindent\textbf{Design Implication.}
Behavior-preserving temporalization should be treated as a first-class systems design decision rather than a fixed preprocessing step. Once coherent behavioral context is preserved, compact transformer models can accurately model multi-stage attacks without substantially increasing model capacity.

We next ask whether preserving behavioral context still requires larger transformers.

\subsection{RQ2: Behavioral Context over Model Scale} 
\label{subsec:finding2}
\noindent\textbf{Finding.} Once behavioral context is preserved through session-coherent temporalization, increasing transformer capacity yields rapidly diminishing returns. Moderate-capacity models achieve detection performance comparable
to that of substantially larger architectures while sustaining significantly higher inference throughput.

\begin{figure}[!ht]
\centering
\vspace{-0.25cm}
\includegraphics[width=0.45\textwidth]{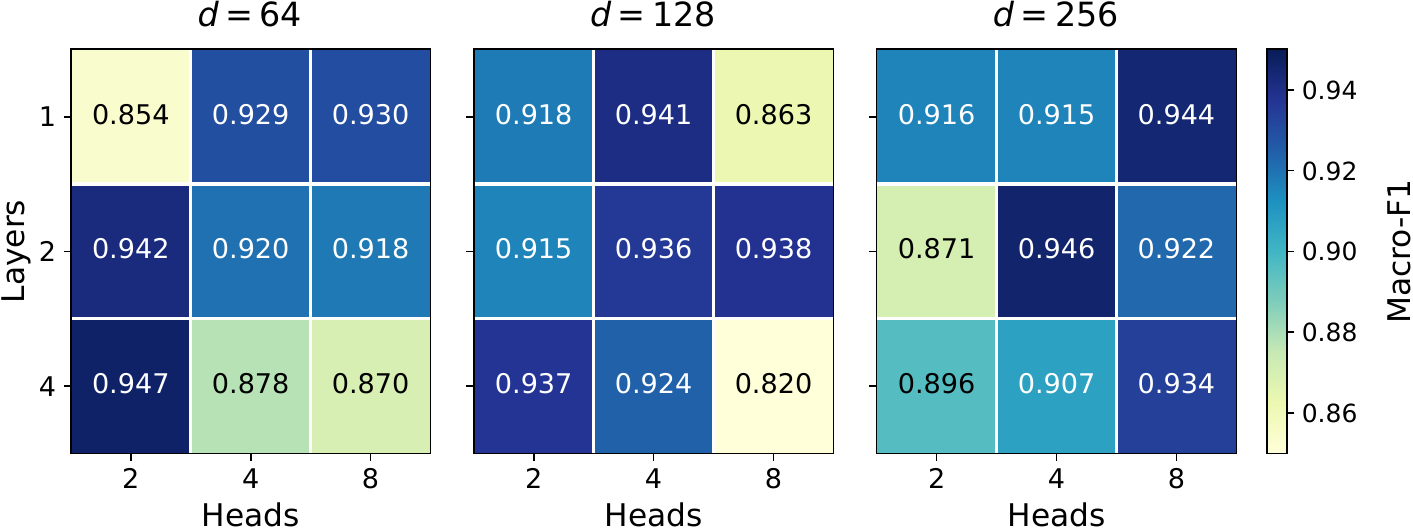}
\vspace{-0.25cm}
\caption{\scriptsize Impact of transformer capacity under session-coherent temporalization.}

\label{fig:capacity_heatmap}
\vspace{-0.35cm}
\end{figure}

Fig.~\ref{fig:capacity_heatmap} evaluates transformer capacity under identical session-coherent temporalization across a grid of embedding dimensions, transformer depths, and attention heads. Because temporalization remains fixed, the experiment isolates the contribution of model capacity. Detection performance improves rapidly when moving from very small models to moderate-capacity configurations, but quickly saturates thereafter despite continued increases in model complexity.

The strongest operating point occurs at $(d,L,H)=(64,4,2)$ with only 0.65M trainable parameters. Scaling to the largest evaluated configuration, $(256,4,8)$ with 3.38M parameters, increases model size by 5.2$\times$ while reducing processed-flow throughput from 0.40M to 0.13M flows/s (3.1$\times$ lower) and slightly decreasing Macro-F1 from 0.947 to 0.934.

These results show that once coherent behavioral context is available, larger transformers provide negligible detection gains while substantially increasing computational cost. These findings are consistent with the hypothesis that preserving behavioral context fundamentally changes the scaling characteristics of transformer-based intrusion detection. By exposing the communication dependencies required for accurate intrusion detection before learning begins, session-coherent temporalization leaves relatively little additional information for larger transformer architectures to exploit.

\begin{figure}[!th]
\centering
\vspace{-0.15cm}
\includegraphics[width=0.90\columnwidth]{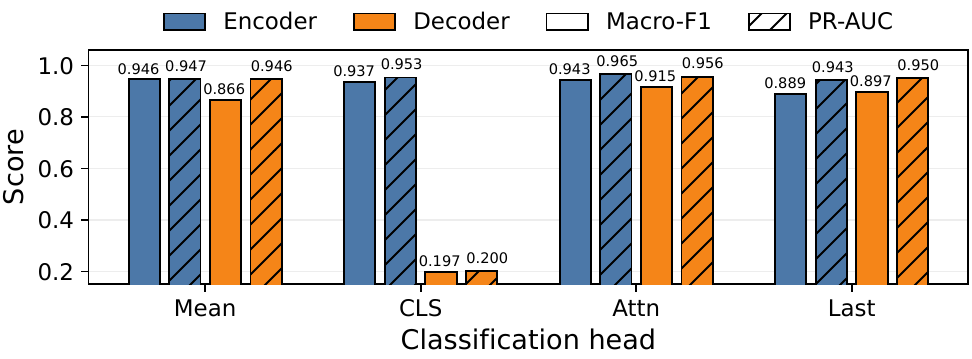}
\vspace{-1.15em}
\caption{\scriptsize Prediction aggregation under fixed transformer capacity $(d{=}256,L{=}2,H{=}4)$. Mean pooling leads for the encoder and attention pooling for the decoder; decoder CLS degrades sharply despite high accuracy.}

\label{fig:head_ablation}
\vspace{-0.30cm}
\end{figure}

\noindent\textbf{Aggregation Has a Secondary Effect.} 
To determine whether prediction aggregation influences this conclusion, we fix the transformer capacity and vary only the sequence aggregation strategy. Fig.~\ref{fig:head_ablation} and Table~\ref{tab:head_ablation} show that mean pooling is strongest for the encoder, whereas attention pooling is strongest for the decoder. Decoder CLS pooling performs substantially worse, reducing Macro-F1 to 0.197 despite maintaining 97\% overall accuracy, highlighting the importance of balanced evaluation metrics for highly imbalanced intrusion detection.

\begin{table}[t]
\caption{\scriptsize Classification head ablation for from-scratch shallow transformer encoder- and decoder-style models under fixed capacity $(d,L,H)=(256,2,4)$. Throughput is reported in flows/s. }
\vspace{-0.3cm}
\label{tab:head_ablation}
\centering
\setlength{\tabcolsep}{1pt}
\begin{tabular}{lllcccc}
\toprule
Model & Family & Head & Macro-F1 & PR-AUC & Accuracy & Flows/s \\
\midrule
ST-Enc & Encoder & Mean & \textbf{0.946} & 0.947 & \textbf{0.996} & 293k \\
ST-Enc & Encoder & CLS  & 0.937 & 0.953 & 0.996 & \textbf{296k} \\
ST-Enc & Encoder & Attn & 0.943 & \textbf{0.965} & \textbf{0.996} & 288k \\
ST-Enc & Encoder & Last & 0.889 & 0.943 & 0.995 & 294k \\
\midrule
ST-Dec &Decoder & Mean & 0.866 & 0.946 & \textbf{0.996} & 251k \\
ST-Dec &Decoder & CLS  & 0.197 & 0.200 & 0.970 & 253k \\
ST-Dec &Decoder & Attn & 0.915 & 0.956 & \textbf{0.996} & 249k \\
ST-Dec &Decoder & Last & 0.897 & 0.950 & \textbf{0.996} & 253k \\
\bottomrule
\end{tabular}
\vspace{-2.25em}
\end{table}

Importantly, the differences between aggregation strategies remain considerably smaller than those observed across temporalization strategies ($\S$\ref{subsec:finding1}). Together, the capacity and aggregation studies demonstrate that once behavioral context is preserved, moderate-capacity transformers are sufficient to model the remaining attack dependencies efficiently.

\noindent\textbf{Design Implication.}
Once behavior-preserving temporalization is in place, increasing transformer capacity offers limited practical benefit. Moderate-capacity transformer models achieve nearly identical detection quality while sustaining substantially higher throughput and lower computational cost, making them a more attractive operating point for deployment.

We next compare the optimized shallow pipeline with substantially larger pretrained models.

\subsection{RQ3: Pipeline Optimization versus Foundation Models}
\label{subsec:finding3}
\noindent\textbf{Finding.} Optimizing the intrusion detection pipeline contributes substantially more to detection performance than increasing model scale. An optimized shallow transformer operating on behavior-preserving input sequences matches or exceeds the performance of considerably larger pretrained foundation models while sustaining one to two orders of magnitude higher inference throughput.

\begin{figure}[!ht]
    \centering
    \vspace{-0.15cm}
\includegraphics[width=0.65\columnwidth]{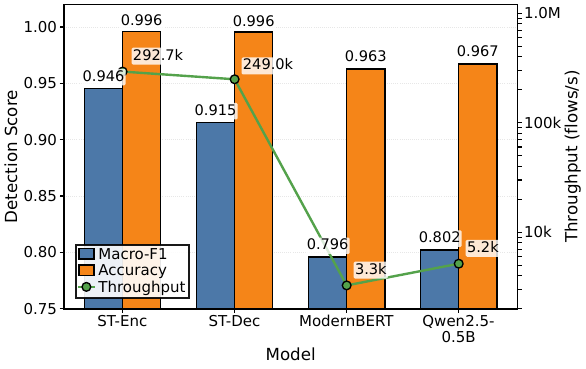}
\vspace{-0.3cm}
     \caption{\scriptsize Comparison between the optimized shallow transformers and representative pretrained transformer models.}
    \label{fig:llm_vs_shallow}
    \vspace{-1.25em}
\end{figure}

Fig.~\ref{fig:llm_vs_shallow} compares from-scratch shallow transformers with representative pretrained foundation models under the same session-coherent temporalization strategy. The shallow baselines are denoted ST-Enc and ST-Dec, while ModernBERT and Qwen2.5-0.5B represent pretrained encoder- and decoder-based foundation models, respectively. Additional pretrained variants, including DeBERTa-v3, CTI-BERT, and ModernBERT-DAPT, are evaluated in the supplementary ablations and artifact~\cite{artifact2027}.

This experiment isolates the effect of model scale under an identical behavior-preserving pipeline. Its objective is to determine whether substantially larger pretrained foundation models provide additional benefit once behavioral context
has already been preserved, rather than to compare their general capabilities across unrelated tasks. The comparison shows that, once communication-session semantics are preserved and the remaining pipeline components are optimized, compact transformers achieve approximately 0.95 Macro-F1 while sustaining inference throughput approaching $10^{6}$ processed flow records per second. In contrast, much larger pretrained models provide little additional balanced-detection gain at substantially greater computational cost.

\begin{table}[t]
\centering
\caption{\scriptsize Shallow and pretrained transformers 
under session-coherent temporalization. ST-B1--ST-B4 are from-scratch shallow Transformer configurations selected from the capacity grid; ModernBERT and Qwen2.5-0.5B are pretrained foundation-model baselines used in Fig.~ \ref{fig:llm_vs_shallow}.}
\label{tab:capacity_summary}
\vspace{-0.3cm}
\scriptsize
\setlength{\tabcolsep}{1.2pt}

\begin{tabular}{llccccc}
\toprule
Model & Config. $(d,L,H)$ & Params & Macro-F1 & PR-AUC & Accuracy & Flows/s \\
\midrule
ST-B1 & $(64,1,2)$ & 0.20M & 0.854 & 0.856 & 0.995 & 1.39M \\
ST-B2 & $(128,1,4)$ & 0.44M & 0.941 & 0.938 & 0.996 & 0.89M \\
ST-B3 & $(64,4,2)$ & 0.65M & 0.947 & 0.958 & 0.996 & 0.40M \\
ST-B4 & $(256,4,8)$ & 3.38M & 0.934 & 0.936 & 0.995 & 0.13M \\
\midrule
ModernBERT & $(768,22,12)$ & 149M & 0.796 & -- & 0.963 & 3.3k \\
Qwen2.5-0.5B & $(896,24,14_{\mathrm{Q}}/2_{\mathrm{KV}})$ & 494M & 0.802 & -- & 0.967 & 5.2k \\
\bottomrule
\end{tabular}

\vspace{0.02cm}
\begin{minipage}{0.98\columnwidth}
\scriptsize
\emph{Note:} For ST rows, $(d,L,H)$ denotes hidden size, depth, and attention heads. For Qwen2.5-0.5B, $14_{\mathrm{Q}}/2_{\mathrm{KV}}$ denotes grouped-query attention. PR-AUC was not part of the seed-averaged pretrained comparison.
\end{minipage}
\vspace{-2.15em}
\end{table}

Table~\ref{tab:capacity_summary} highlights the substantial difference in model scale. ModernBERT (149M parameters) and Qwen2.5-0.5B (494M) are approximately $229\times$ and $759\times$ larger than the strongest shallow operating point (ST-B3, 0.65M), respectively. Despite this difference, neither pretrained model improves Macro-F1, while inference throughput decreases by up to two orders of magnitude. 

This result highlights an important distinction between natural language processing and flow-based intrusion detection. Large language models derive much of their advantage from modeling rich linguistic structure over long textual contexts. In contrast, network traffic consists of structured communication events whose usefulness depends primarily on preserving interaction semantics during sequence construction. With coherent behavioral context, the remaining task becomes simpler,
reducing the need for larger models.

These findings do not imply that foundation models are
ineffective. Rather, under the evaluated flow-based NIDS
pipeline, preserving communication context provides
greater practical benefit than increasing model capacity.

Collectively, these results shift the systems design emphasis from model scaling toward pipeline engineering. Rather than asking \emph{``How can we build a larger transformer?''}, the more important question for transformer-based intrusion detection becomes \emph{``How should network traffic be transformed before any transformer is applied?''}

\noindent\textbf{Design Implication.} Future transformer-based NIDS should prioritize behavior-preserving pipeline design before increasing model capacity. Once communication-session semantics are preserved, lightweight transformers balance detection quality, efficiency, and throughput, making them better suited to high-speed deployment than much larger
foundation models.

\subsection{RQ4: Deployment Trade-offs}
\label{subsec:finding4}
Having shown that pipeline optimization consistently outperforms model scaling, we next examine whether these gains translate into practical deployment advantages.

\noindent\textbf{Finding.}
Optimizing sequence construction enables compact transformer models to achieve an attractive deployment operating point, simultaneously providing high detection quality, high throughput, and robust performance across diverse datasets. Larger pretrained models offer limited detection gains at substantially higher cost.

\begin{figure}[t]
    \centering
    \includegraphics[width=0.66\columnwidth]{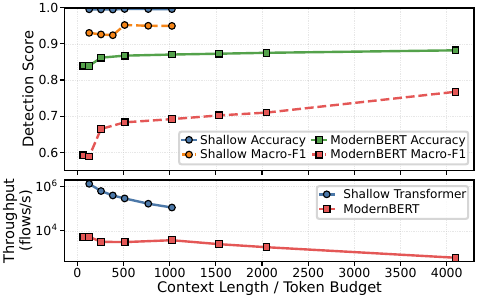}
    \vspace{-0.8em}
    \caption{\scriptsize Effect of context length on detection quality and throughput for shallow transformers and ModernBERT. }
    \label{fig:context_tradeoff}
    \vspace{-1.15em}
\end{figure}

\begin{figure}[t]
    \centering
    \includegraphics[width=0.62\columnwidth]{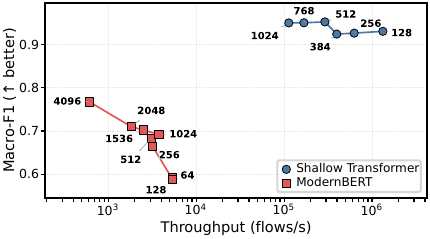}
    
    \caption{\scriptsize Efficiency--performance frontier for shallow transformers and ModernBERT.}
    \label{fig:pareto_tradeoff}
    \vspace{-1.25em}
\end{figure}

Fig.~\ref{fig:context_tradeoff} shows that longer contexts gradually improve ModernBERT Macro-F1 but sharply reduce throughput. By contrast, the shallow transformer maintains near-constant detection quality and much higher throughput, indicating that session-coherent temporalization supplies sufficient context without longer inputs. 
Fig.~\ref{fig:pareto_tradeoff} depicts the deployment trade-off. Shallow transformers occupy the upper-right region of the Pareto frontier, achieving Macro-F1 values approaching 0.95 while sustaining approximately $10^{5}$--$10^{6}$ processed flow records per second. In contrast, pretrained foundation models move along the frontier toward substantially higher computational cost without providing proportional improvements in detection quality.

\begin{figure}[!t]
    \centering
    \includegraphics[width=0.65\columnwidth]
    {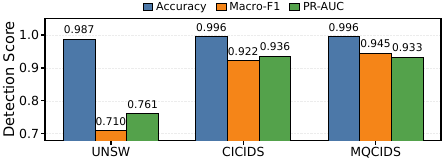}
    \caption{\scriptsize Per-dataset validation of the selected transformer-based intrusion
    detection configuration across the three evaluated datasets.}
    \label{fig:dataset_validation}
    \vspace{-2.0em}
\end{figure}

\noindent{\bf Per-dataset Validation.} Deployment efficiency alone is insufficient if performance depends on a particular benchmark. We therefore evaluate the deployment configuration selected from the preceding efficiency--accuracy analysis across all three datasets. Fig.~\ref{fig:dataset_validation} shows that the shallow-transformer model performs strongly across UNSW, CICIDS, and MQCIDS. Accuracy remains high (0.987--0.996), while Macro-F1 reaches 0.710, 0.922, and 0.945, with corresponding PR-AUC values of 0.761, 0.936, and 0.933. The lower performance on UNSW reflects its greater class imbalance and attack diversity, whereas CICIDS and MQCIDS achieve consistently strong balanced detection.

Collectively, these experiments demonstrate that session-coherent temporalization fundamentally improves the accuracy--efficiency trade-off of transformer-based NIDS. By preserving communication-session semantics, lightweight transformers perform strongly across all three benchmarks at substantially lower inference cost than larger models; separately,
session-coherent window construction sustains 1.04 million input flows/s despite measurable overhead. These findings suggest that improving sequence construction is a more effective strategy for advancing NIDS performance than simply increasing model complexity.

\noindent\textbf{Design Recommendation.}

Future transformer-based NIDS should prioritize behavior-preserving sequence construction before increasing model capacity.

\subsection{RQ5: Cross-Dataset Generalization}
\label{subsec:cross_dataset}
Having established the deployment advantages of the proposed pipeline, we finally evaluate whether its benefits generalize beyond the datasets on which it was trained.

\noindent\textbf{Finding.} Cross-dataset generalization is primarily limited by domain shift rather than sequence construction. Session-coherent temporalization provides modest average improvements in zero-shot transfer, although heterogeneous traffic distributions remain the dominant challenge.

Unlike the previous within-dataset evaluations, this experiment performs zero-shot cross-dataset transfer, where models are trained on one dataset and evaluated directly on another without fine-tuning or domain adaptation. As shown in Table~\ref{tab:cross_dataset}, session-coherent temporalization increases the average Macro-F1 from 0.414 to 0.423 and reduces the average TRG from 0.429 to 0.419, with improvement or parity in five of the six transfer directions. The largest gains occur for UNSW$\rightarrow$MQCIDS (+0.056 Macro-F1) and UNSW$\rightarrow$CICIDS (+0.026), indicating that preserving communication-session semantics improves behavioral transfer under moderate domain shift. Across all transfer tasks, an exact one-sided Wilcoxon signed-rank test indicates positive but statistically inconclusive trends for both Macro-F1 ($W^{+}=11$, $p=0.19$, $r_{\mathrm{rb}}=0.47$) and TRG ($W^{+}=14$, $p=0.28$, $r_{\mathrm{rb}}=0.33$).

These gains are smaller than those observed in RQ1 because zero-shot transfer must additionally generalize across different traffic distributions, attack compositions, feature characteristics, and measurement environments. Consequently, domain shift dominates the remaining performance degradation, limiting the impact of improved sequence construction alone. This is illustrated by the reduced performance for MQCIDS$\rightarrow$CICIDS, where substantial distribution differences outweigh the benefits of preserving communication-session semantics. Overall, session-coherent temporalization improves transfer robustness but does not eliminate the need for target-domain validation or lightweight adaptation when deploying across heterogeneous environments.

\begin{table}[!t]
\centering
\scriptsize
\caption{Behavior preservation under domain shift. Models are trained
on one dataset and evaluated directly on another without fine-tuning.
$\Delta$F1 is Session minus Sliding; positive $\Delta$TRG indicates a smaller
TRG for Session. }

\vspace{-1.15em}
\label{tab:cross_dataset}
\setlength{\tabcolsep}{2.1pt}
\begin{tabular}{llrrrrrr}
\toprule
Train & Test & Slide. F1 & Sess. F1 & $\Delta$F1 &
TRG$_{\mathrm{Slide}}$ & TRG$_{\mathrm{Sess}}$ & $\Delta$TRG \\
\midrule
CICIDS & UNSW   & 0.420 & 0.423 &  0.003 & 0.199 & 0.198 &  0.001 \\
CICIDS & MQCIDS & 0.333 & 0.333 &  0.000 & 0.181 & 0.161 &  0.020 \\
UNSW   & MQCIDS & 0.383 & 0.439 &  0.056 & 0.612 & 0.557 &  0.055 \\
UNSW   & CICIDS & 0.480 & 0.506 &  0.026 & 0.477 & 0.451 &  0.026 \\
MQCIDS & UNSW   & 0.479 & 0.482 &  0.003 & 0.506 & 0.510 & -0.004 \\
MQCIDS & CICIDS & 0.387 & 0.355 & -0.032 & 0.598 & 0.637 & -0.039 \\
\midrule
Avg.   & --     & 0.414 & 0.423 &  0.009 & 0.429 & 0.419 &  0.010 \\
\bottomrule
\end{tabular}
\vspace{-1.75em}
\end{table}

\textbf{Design Implication.}
Session-coherent temporalization provides a strong foundation for transformer-based NIDS, but robust cross-domain deployment also requires online adaptation. Integrating it with recent adaptation frameworks, such as AOC-IDS, which combines contrastive representation learning and pseudo-label-driven updates~\cite{zhang2024aocids}, is a promising direction for improving robustness under domain shift.

\section{Conclusion}
\label{sec:conclusion}
This paper showed that the primary bottleneck in transformer-based intrusion detection is not model capacity, but sequence construction. We introduced \emph{session-coherent temporalization}, a behavior-preserving strategy that organizes flows according to communication sessions, enabling lightweight transformers to consistently match or outperform substantially larger pretrained models while achieving substantially higher deployment efficiency. Although cross-dataset transfer remains affected by domain shift, the results demonstrate that preserving behavioral context is a fundamental systems design principle for transformer-based NIDS. Future work will investigate online contrastive adaptation over session-coherent flow sequences to improve robustness under evolving traffic distributions and previously unseen environments. More broadly, our findings suggest that future advances in sequential learning will depend as much on constructing meaningful input sequences as on developing increasingly larger learning architectures.

\balance
\bibliographystyle{IEEEtranS}
\bibliography{sample}

@inproceedings{roesch1999snort,
  author    = {Roesch, Martin},
  title     = {Snort: Lightweight Intrusion Detection for Networks},
  booktitle = {Proceedings of the 13th USENIX Conference on System Administration (LISA)},
  year      = {1999},
  pages     = {229--238}
}

@misc{artifact2027,
  author = {Shah, Zeeshan Zulkifl and
            Ikram, Muhammad and
            Kaafar, Dali and
            Zhou, Yipeng},
  title = {{Beyond Model Scaling: Rethinking Sequence
           Construction for Transformer-based Network
           Intrusion Detection} -- Reproducibility Artifact},
  year = {2026},
  howpublished = {\url{https://github.com/ZeeshanZulkiflShah/sequence-construction-nids-artifact/tree/arxiv-v1-artifact}},
  note = {Versioned reproducibility artifact}
}

@inproceedings{corsini2021evaluation,
  title={On the evaluation of sequential machine learning for network intrusion detection},
  author={Corsini, Andrea and Yang, Shanchieh Jay and Apruzzese, Giovanni},
  booktitle={Proceedings of the 16th International Conference on Availability, Reliability and Security},
  pages={1--10},
  year={2021}
}

@inproceedings{shen2019attack2vec,
  title={{ATTACK2VEC}: Leveraging temporal word embeddings to understand the evolution of cyberattacks},
  author={Shen, Yun and Stringhini, Gianluca},
  booktitle={28th USENIX Security Symposium (USENIX Security 19)},
  pages={905--921},
  year={2019}
}

@inproceedings{sommer2010outside,
  author={Sommer, Robin and Paxson, Vern},
  booktitle={2010 IEEE Symposium on Security and Privacy}, 
  title={Outside the Closed World: On Using Machine Learning for Network Intrusion Detection}, 
  year={2010},
  volume={},
  number={},
  pages={305-316},
  doi={10.1109/SP.2010.25}}

@article{buczak2016survey,
author = {Buczak, Anna L. and Guven, Erhan},
title = {A Survey of Data Mining and Machine Learning Methods for Cyber Security Intrusion Detection},
year = {2016},
issue_date = {Secondquarter 2016},
publisher = {IEEE Press},
volume = {18},
number = {2},
issn = {1553-877X},
doi = {10.1109/COMST.2015.2494502},
journal = {Commun. Surveys Tuts.},
month = apr,
pages = {1153--1176},
numpages = {24}
}

@inproceedings{park_you_2023_ctibert,
  author    = {Park, Youngja and You, Weiqiu},
  title     = {A Pretrained Language Model for Cyber Threat Intelligence},
  booktitle = {Proceedings of the 2023 Conference on Empirical Methods
               in Natural Language Processing: Industry Track},
  year      = {2023},
  pages     = {113--122},
  publisher = {Association for Computational Linguistics},
  doi       = {10.18653/v1/2023.emnlp-industry.12}
}

@article{martin2026transformer,
  author  = {Martin-Reizabal, Sergio and Caballero-Quiroga, Adrian
             and Gil-Arroyo, Beatriz and Basurto, Nu{\~n}o
             and Ruiz-Gonzalez, Ruben},
  title   = {Transformer-Based Classification of {IoT} Network Traffic
             with Flow-to-Window Aggregation},
  journal = {Internet of Things},
  volume  = {36},
  pages   = {101879},
  year    = {2026},
  doi     = {10.1016/j.iot.2026.101879}
}

@inproceedings{vaswani2017attention,
 author = {Vaswani, Ashish and Shazeer, Noam and Parmar, Niki and Uszkoreit, Jakob and Jones, Llion and Gomez, Aidan N and Kaiser, \L ukasz and Polosukhin, Illia},
 booktitle = {Advances in Neural Information Processing Systems},
 editor = {I. Guyon and U. Von Luxburg and S. Bengio and H. Wallach and R. Fergus and S. Vishwanathan and R. Garnett},
 pages = {},
 publisher = {Curran Associates, Inc.},
 title = {Attention is All you Need},
 url = {https://proceedings.neurips.cc/paper_files/paper/2017/file/3f5ee243547dee91fbd053c1c4a845aa-Paper.pdf},
 volume = {30},
 year = {2017}
}

@misc{paszke2019pytorch,
      title={{PyTorch}: An Imperative Style, High-Performance Deep Learning Library}, 
      author={Adam Paszke and Sam Gross and Francisco Massa and Adam Lerer and James Bradbury and Gregory Chanan and Trevor Killeen and Zeming Lin and Natalia Gimelshein and Luca Antiga and Alban Desmaison and Andreas Köpf and Edward Yang and Zach DeVito and Martin Raison and Alykhan Tejani and Sasank Chilamkurthy and Benoit Steiner and Lu Fang and Junjie Bai and Soumith Chintala},
      year={2019},
      eprint={1912.01703},
      archivePrefix={arXiv},
      primaryClass={cs.LG},
      url={https://arxiv.org/abs/1912.01703}, 
}

@article{manocchio2024flowtransformer,
  author  = {Manocchio, Liam Daly and Layeghy, Siamak and Lo, Wai Weng and Kulatilleke, Gayan K. and Sarhan, Mohanad and Portmann, Marius},
  title   = {{FlowTransformer}: A Transformer Framework for Flow-Based Network Intrusion Detection Systems},
  journal = {Expert Systems with Applications},
  volume  = {241},
  pages   = {122564},
  year    = {2024},
  doi     = {10.1016/j.eswa.2023.122564}
}

@inproceedings{moustafa2015unsw,
  author    = {Moustafa, Nour and Slay, Jill},
  title     = {{UNSW-NB15}: A Comprehensive Data Set for Network Intrusion Detection Systems ({UNSW-NB15} Network Data Set)},
  booktitle = {2015 Military Communications and Information Systems Conference (MilCIS)},
  year      = {2015},
  pages     = {1--6},
  doi       = {10.1109/MilCIS.2015.7348942}
}

@inproceedings{sharafaldin2018toward,
  author    = {Sharafaldin, Iman and Habibi Lashkari, Arash
               and Ghorbani, Ali A.},
  title     = {Toward Generating a New Intrusion Detection Dataset
               and Intrusion Traffic Characterization},
  booktitle = {Proceedings of the 4th International Conference on
               Information Systems Security and Privacy (ICISSP)},
  year      = {2018},
  pages     = {108--116},
  doi       = {10.5220/0006639801080116}
}

@misc{he2023debertav3,
      title={{DeBERTaV3}: Improving {DeBERTa} using {ELECTRA}-Style Pre-Training with Gradient-Disentangled Embedding Sharing}, 
      author={Pengcheng He and Jianfeng Gao and Weizhu Chen},
      year={2023},
      eprint={2111.09543},
      archivePrefix={arXiv},
      primaryClass={cs.CL},
      url={https://arxiv.org/abs/2111.09543}, 
}

@misc{warner2025modernbert,
      title={Smarter, Better, Faster, Longer: A Modern Bidirectional Encoder for Fast, Memory Efficient, and Long Context Finetuning and Inference}, 
      author={Benjamin Warner and Antoine Chaffin and Benjamin Clavié and Orion Weller and Oskar Hallström and Said Taghadouini and Alexis Gallagher and Raja Biswas and Faisal Ladhak and Tom Aarsen and Nathan Cooper and Griffin Adams and Jeremy Howard and Iacopo Poli},
      year={2024},
      eprint={2412.13663},
      archivePrefix={arXiv},
      primaryClass={cs.CL},
      url={https://arxiv.org/abs/2412.13663}, 
}

@misc{qwen2025qwen25technicalreport,
  author        = {Yang, An and others},
  title         = {{Qwen2.5 Technical Report}},
  year          = {2024},
  eprint        = {2412.15115},
  archiveprefix = {arXiv},
  primaryclass  = {cs.CL},
  doi           = {10.48550/arXiv.2412.15115},
  url           = {https://arxiv.org/abs/2412.15115}
}

@article{hutchins2011killchain,
  title={Intelligence-driven computer network defense informed by analysis of adversary campaigns and intrusion kill chains},
  author={Hutchins, Eric M and Cloppert, Michael J and Amin, Rohan M and others},
  journal={Leading Issues in Information Warfare \& Security Research},
  volume={1},
  number={1},
  pages={80},
  year={2011}
}

@article{chou2022csur,
author = {Chou, Dylan and Jiang, Meng},
title = {A Survey on Data-driven Network Intrusion Detection},
year = {2021},
issue_date = {December 2022},
publisher = {Association for Computing Machinery},
address = {New York, NY, USA},
volume = {54},
number = {9},
issn = {0360-0300},
doi = {10.1145/3472753},
journal = {ACM Comput. Surv.},
articleno = {182},
numpages = {36}
}

@misc{suricata_doc,
  author = {{Suricata}},
  title  = {Suricata User Guide — Suricata 8.0.5 documentation},
  year   = {2025},
  url    = {https://docs.suricata.io/}
}

@misc{snort_docs,
  author       = {{Snort Project}},
  title        = {Snort 3 Rule Writing Guide and Documentation},
  year         = {2026},
  url          = {https://docs.snort.org/},
  note   = {Accessed 2026-03-19}
}

@misc{google_https,
  author = {{Google}},
  title  = {Google Transparency Report: {HTTPS} Encryption on the Web},
  year   = {2026},
  url    = {https://transparencyreport.google.com/https/overview},
  note   = {Accessed 2026-03-19}
}

@inproceedings{felt2017https,
  author    = {Felt, Adrienne Porter and Barnes, Richard and King, April and Palmer, Chris and Bentzel, Chris and Tabriz, Parisa},
  title     = {Measuring {HTTPS} Adoption on the Web},
  booktitle = {Proceedings of the 26th USENIX Security Symposium (USENIX Security)},
  year      = {2017},
  pages     = {1323--1338}
}

@inproceedings{wolf-etal-2020-transformers,
    title = "Transformers: State-of-the-Art Natural Language Processing",
    author = "Thomas Wolf and Lysandre Debut and Victor Sanh and Julien Chaumond and Clement Delangue and Anthony Moi and Pierric Cistac and Tim Rault and Rémi Louf and Morgan Funtowicz and Joe Davison and Sam Shleifer and Patrick von Platen and Clara Ma and Yacine Jernite and Julien Plu and Canwen Xu and Teven Le Scao and Sylvain Gugger and Mariama Drame and Quentin Lhoest and Alexander M. Rush",
    booktitle = "Proceedings of the 2020 Conference on Empirical Methods in Natural Language Processing: System Demonstrations",
    month = oct,
    year = "2020",
    address = "Online",
    publisher = "Association for Computational Linguistics",
    url = "https://aclanthology.org/2020.emnlp-demos.6/",
    pages = "38--45"
}

@inproceedings{pr-auc,
author = {Davis, Jesse and Goadrich, Mark},
title = {The relationship between Precision-Recall and {ROC} curves},
year = {2006},
isbn = {1595933832},
publisher = {Association for Computing Machinery},
address = {New York, NY, USA},
url = {https://doi.org/10.1145/1143844.1143874},
doi = {10.1145/1143844.1143874},
booktitle = {Proceedings of the 23rd International Conference on Machine Learning},
pages = {233--240},
numpages = {8},
location = {Pittsburgh, Pennsylvania, USA},
series = {ICML '06}
}

@article{han2023gtid,
title = {Network intrusion detection based on n-gram frequency and time-aware transformer},
journal = {Computers \& Security},
volume = {128},
pages = {103171},
year = {2023},
issn = {0167-4048},
doi = {10.1016/j.cose.2023.103171},
author = {Xueying Han and Susu Cui and Song Liu and Chen Zhang and Bo Jiang and Zhigang Lu}
}

@inproceedings{shah2025mqcids,
  author    = {Shah, Zeeshan Zulkifl and Ikram, Muhammad and Asghar, Hassan Jameel and Kaafar, Mohamed Ali},
  title     = {Deception Meets Diagnostics: Deception-Based Real-Time Threat Detection in Healthcare Web Systems},
  booktitle = {Proceedings of the 28th International Symposium on Research in Attacks, Intrusions and Defenses (RAID)},
  year      = {2025},
  pages     = {391--410},
  publisher = {IEEE Computer Society},
  doi       = {10.1109/RAID67961.2025.00060}
}

@misc{ibmresearch_ctibert_hf,
  author = {{IBM Research}},
  title  = {{ibm-research/CTI-BERT Hugging Face}},
  year   = {2025},
  url    = {https://huggingface.co/ibm-research/CTI-BERT}
}

@inproceedings{liu2019fsnet,
  author    = {Chang Liu and Longtao He and Gang Xiong and Zigang Cao
               and Zhen Li},
  title     = {{FS-Net}: A Flow Sequence Network for Encrypted Traffic
               Classification},
  booktitle = {IEEE INFOCOM 2019 - IEEE Conference on Computer
               Communications},
  pages     = {1171--1179},
  year      = {2019},
  doi       = {10.1109/INFOCOM.2019.8737507}
}

@inproceedings{diallo2021acid,
  author    = {Alec F. Diallo and Paul Patras},
  title     = {Adaptive Clustering-Based Malicious Traffic
               Classification at the Network Edge},
  booktitle = {IEEE INFOCOM 2021 - IEEE Conference on Computer
               Communications},
  pages     = {1--10},
  year      = {2021},
  doi       = {10.1109/INFOCOM42981.2021.9488690}
}

@inproceedings{zhang2024aocids,
  author    = {Xinchen Zhang and Running Zhao and Zhihan Jiang and
               Zhicong Sun and Yulong Ding and Edith C. H. Ngai and
               Shuang-Hua Yang},
  title     = {{AOC-IDS}: Autonomous Online Framework with Contrastive
               Learning for Intrusion Detection},
  booktitle = {IEEE INFOCOM 2024 - IEEE Conference on Computer
               Communications},
  pages     = {581--590},
  year      = {2024},
  doi       = {10.1109/INFOCOM52122.2024.10621346}
}

@inproceedings{zhao2024rids,
  author    = {Ziming Zhao and Zhaoxuan Li and Zhuoxue Song and
               Fan Zhang and Binbin Chen},
  title     = {{RIDS}: Towards Advanced {IDS} via {RNN} Model and
               Programmable Switches Co-Designed Approaches},
  booktitle = {IEEE INFOCOM 2024 - IEEE Conference on Computer
               Communications},
  pages     = {591--600},
  year      = {2024},
  doi       = {10.1109/INFOCOM52122.2024.10621290}
}

@inproceedings{lin2022etbert,
  author    = {Xinjie Lin and Gang Xiong and Gaopeng Gou and
               Zhen Li and Junzheng Shi and Jing Yu},
  title     = {{ET-BERT}: A Contextualized Datagram Representation
               with Pre-Training Transformers for Encrypted Traffic
               Classification},
  booktitle = {Proceedings of the ACM Web Conference 2022},
  pages     = {633--642},
  publisher = {ACM},
  year      = {2022},
  doi       = {10.1145/3485447.3512217}
}

@inproceedings{zhao2020pigasus,
  author    = {Zhipeng Zhao and Hugo Sadok and Nirav Atre and
               James C. Hoe and Vyas Sekar and Justine Sherry},
  title     = {Achieving {100Gbps} Intrusion Prevention on a Single Server},
  booktitle = {14th USENIX Symposium on Operating Systems Design
               and Implementation (OSDI 20)},
  pages     = {1083--1100},
  publisher = {USENIX Association},
  year      = {2020}
}

\end{document}